\documentclass[jgrga]{AGUTeX}

\usepackage{graphicx}
\usepackage{url}
\usepackage{rotating}

\authorrunninghead{LUGAZ ET AL.}

\titlerunninghead{GEO-EFFECTIVENESS OF SHOCKS AND SHEATHS}

\authoraddr{Corresponding author: N. Lugaz,
Department of Physics and Space Science Center, University of
New Hampshire, Morse Hall, 8 College Rd, Durham, NH 03824, USA.
(noe.lugaz@unh.edu)}

\begin{document}

%
%

\title{Factors Affecting the Geo-effectiveness of Shocks and Sheaths at 1 AU}

\authors{N.\ Lugaz\altaffilmark{1,2}, C.\ J.~Farrugia\altaffilmark{1,2}, R.\ M.~Winslow\altaffilmark{1}, N.~Al-Haddad\altaffilmark{3, 2},  E.\ K.\ J.~Kilpua \altaffilmark{4}, P. Riley\altaffilmark{5}}

\altaffiltext{1}{Space Science Center, University of New Hampshire, Durham, NH, USA}
\altaffiltext{2}{Department of Physics, University of New Hampshire, Durham, NH, USA}
\altaffiltext{3}{Institute for Astrophysics and Computational Sciences, Catholic University of America, Washington, DC, USA}
\altaffiltext{4}{Department of Physics, University of Helsinki, Helsinki, Finland}
\altaffiltext{5}{Predictive Sciences Inc., San Diego, CA, USA}

%
%

\begin{abstract}
We identify all fast-mode forward shocks, whose sheath regions resulted in a moderate (56 cases) or intense (38 cases) geomagnetic storm during 18.5 years from January 1997 to June 2015. We study their main properties, interplanetary causes and geo-effects.
We find that half (49/94) such shocks are associated with interacting coronal mass ejections (CMEs), as they are either shocks propagating into a preceding CME (35 cases) or a shock propagating into the sheath region of a preceding shock (14 cases). About half (22/45) of the shocks driven by isolated transients and which have geo-effective sheaths compress pre-existing southward $B_z$. Most of the remaining sheaths appear to have planar structures with southward magnetic fields, including some with planar structures consistent with field line draping ahead of the magnetic ejecta.  
A typical (median) geo-effective shock-sheath structure drives a geomagnetic storm with peak Dst of $-88$~nT, pushes the subsolar magnetopause location to 6.3~R$_\mathrm{E}$, i.e.\ below geosynchronous orbit and is associated with substorms with a peak AL-index of $-1350$~nT.  There are some important differences between sheaths associated with CME-CME interaction (stronger storms) and  those associated with isolated CMEs (stronger compression of the magnetosphere). 
We detail six case studies of different types of geo-effective shock-sheaths, as well as two events for which there was no geomagnetic storm but other magnetospheric effects. Finally, we discuss our results in terms of space weather forecasting, and potential effects on Earth's radiation belts.

\end{abstract}

\begin{article}

\section{INTRODUCTION} \label{intro}

Starting in the early 1970s, the relative importance of shocks and magnetic ejecta as the drivers of geomagnetic storms was heavily debated. By the 1980s, it had been recognized that magnetic ejecta, in particular magnetic clouds, are the main drivers of intense geomagnetic activity \citep[]{Burlaga:1982, Burlaga:1987, Gosling:1990,Farrugia:1993b, Gonzalez:1994}. One of the quantities that correlates best with the geo-effectiveness of an interplanetary event is its dawn-to-dusk electric field ($VB_z$, where $V$ is the solar wind speed and $B_z$ the north-south component of the magnetic field vector in the GSM coordinate system). Typically, when $VB_z$ reaches $-3$ mV\,m$^{-1}$ for a duration of three hours or more, a major (defined by the disturbance storm-time --Dst-- index being less than $-100$~nT) geomagnetic storm ensues \citep[see][and references therein]{Gonzalez:1987,Bothmer:2007}. 
Magnetic ejecta, and, in particular their subset magnetic clouds, are characterized by slowly varying and enhanced magnetic fields of long duration ($\sim$ 1 day at 1~AU) \citep[]{Burlaga:1981}. When favorably oriented, they may result in moderate or intense geomagnetic storms, as both the speed and the magnetic field are enhanced \citep[]{Zhang:1988, Gonzalez:1988,Tsurutani:1988, Gosling:1991,Farrugia:1993b}. 

However, magnetic clouds are not the only source of geomagnetic storms: \citet{Tsurutani:1988, Tsurutani:1992} and \citet{Huttunen:2002} showed that  shocks/sheaths may also be geoeffective. Note that, in some cases, only a shock and sheath is observed and the bulk of the ejecta is missed; many of these shocks are nonetheless thought to be driven by CMEs \citep[]{Jian:2006,Janvier:2014, Kilpua:2015}. Another key large-scale interplanetary structure driving geomagnetic storms is a corotating interaction region (CIR) that forms at the interface between slow and fast solar wind streams. CIRs, some of which also drive shocks at 1 AU, produce mainly moderate storms \citep[]{Gosling:1991,Bothmer:2004,Echer:2013}. Lastly, more complex, ``compound'' events that arise from the interaction between multiple CMEs or between CMEs and CIRs in the heliosphere \citep[]{Burlaga:1987,Tsurutani:1988} can also drive geomagnetic storms, often intense ones \citep[]{Bothmer:2004,Bothmer:2007}.

The availability of continuous {\it in situ} measurements with high temporal resolution from near L1 provided by Wind since 1995 and ACE since 1998 allows performing large surveys of the interplanetary causes of geomagnetic storms. For instance, \citet{Zhang:2007} studied all 88 intense geomagnetic storms  
that occurred from 1997 to 2005, \citet{Echer:2008} all 90 intense geomagnetic storms from 1997 to 2006 (the 2 additional storms occurred in 2006). \citet{Yermolaev:2010} and \citet{Echer:2013} studied moderate storms (minimum Dst below $-50$~nT but greater than $-100$~nT) during the same time period. These studies found that 25--35\% of intense geomagnetic storms and 15--25\% of moderate geomagnetic storms are caused by the sheath plasma and magnetic field behind fast forward magnetosonic shocks, confirming previous studies such as those of \citet{Gonzalez:1987}, \citet{Gosling:1991} and \citet{Bothmer:2007}. 
Overall, and based on the Dst index only, shocks and their sheaths (hereafter called
shock-sheaths) are found to be second only to magnetic clouds for inducing intense geomagnetic effects, more important than heliospheric current sheet crossings, or  CIRs. Other studies, such as one focusing on 56 intense geomagnetic storms from 1997 to 2002 by \citet{Huttunen:2004} found that shocks-sheaths caused the largest fraction of storms ($\sim$~45\%), and the result may depend on the phase of the solar cycle that is studied.

Most fast forward shocks measured near 1 AU propagate into a typical, unperturbed slow solar wind, but some may also propagate within a preceding CME or magnetic cloud. The possibility that a shock overtaking a magnetic ejecta may result in intense geomagnetic activities was raised in \citet{Ivanov:1982} and \citet{Burlaga:1987} and discussed in greater detail in \citet{Wang:2003a}. Dedicated studies of {\it in situ} measurements of shocks propagating within a preceding magnetic ejecta were reported in \citet{Collier:2007} and \citet{Lugaz:2015a}; an example of their interplanetary formation was discussed in \citet{Liu:2014b}; an example of their potential effects on the radiation belt has been described in \citet{Lugaz:2015b}; and simulations of such events was performed in \citet{Vandas:1997}, \citet{Lugaz:2005b}, \citet{Xiong:2006} and \citet{Lugaz:2013b}. This type of event can be thought as an ongoing CME-CME interaction event.

Shocks propagating within magnetic ejecta were also considered as a separate category in many studies looking at the interplanetary causes of geomagnetic storms: \citet{Zhang:2007} listed ``preceding CME - shock combination'' as the cause of nine out of their 88 intense storms (see more statistics in section~2). \citet{Lugaz:2015a} noted that 19 of their 59 studied events resulted in a drop of Dst of at least $-50$~nT within nine hours of shock arrival. In addition,   \citet{Zhang:2007} identified series of fast forward magnetosonic shocks as a potential cause of geomagnetic storms. These shocks can be considered to be an extension of the set of shocks propagating within a magnetic ejecta since they are typically shocks propagating within a sheath region of a preceding shock. \citet{Zhang:2007} found two such series of shocks causing an intense geomagnetic storm, while nine others were part of the sequence of events resulting in an intense geomagnetic storm (for example a series of shocks, followed by a magnetic ejecta or by a series of magnetic ejecta).

\begin{figure*}[tb]
\centering
{\includegraphics*[width=6.5in]{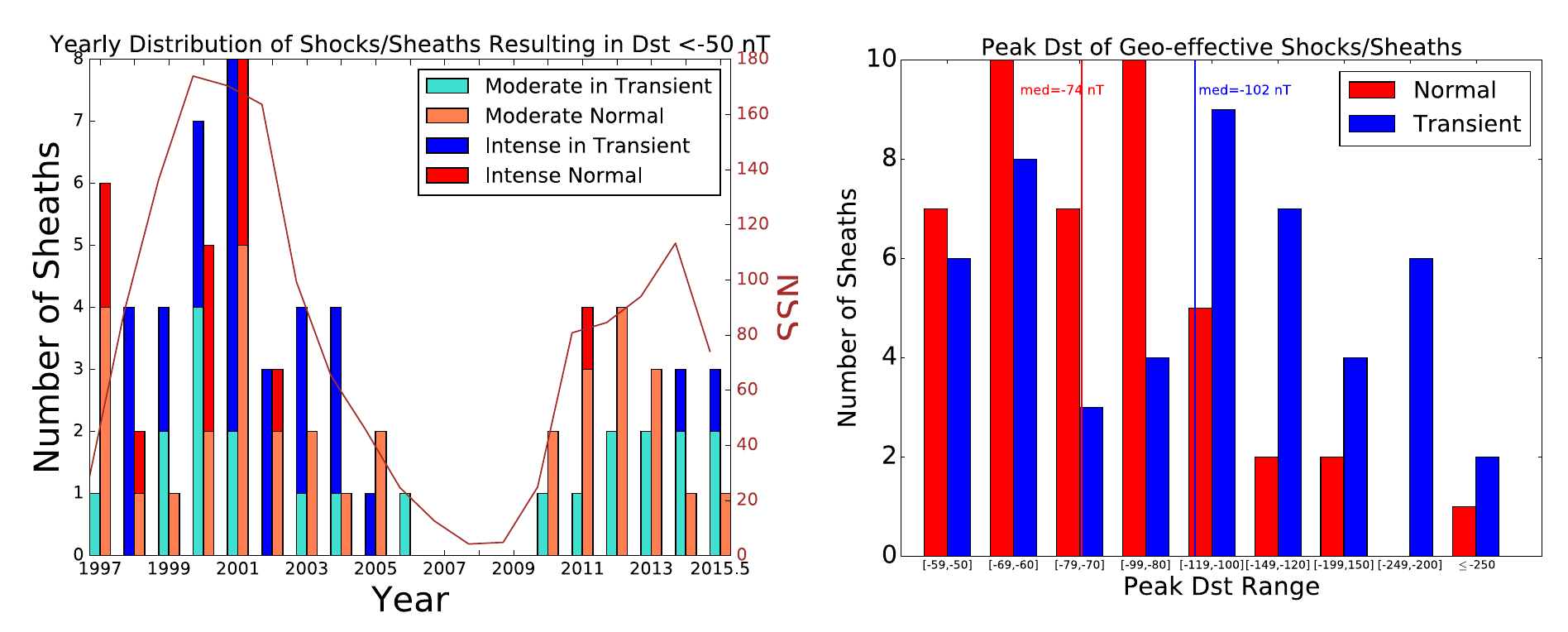}}
\caption{Left: Yearly distribution of shocks for which the sheath produces at least a moderate geomagnetic storm separated into shocks propagating into nominal solar wind conditions (``Normal'') in salmon and red, and shocks propagating into a preceding transient (``in Transient'') in turquoise and blue. The yearly sunspot number is shown in brown on the right-hand scale. Right: Distribution of peak Dst in the sheaths resulting in geomagnetic storms. The median value of the peak Dst for geo-effective sheaths is shown with the vertical lines.}
\label{fig:dist}
\end{figure*}

The reasons why some, but not all, CMEs are geo-effective have been well studied; the same cannot be said for shock-sheaths. This is an important issue as there were over 300 fast forward shocks that impacted Earth during solar cycle (SC) 23 (following the database of \url{http://ipshocks.fi}), and only about 10\% resulted in an intense geomagnetic storm. For magnetic clouds, on the other hand, about one third (31/90 during the same time period) resulted in an intense geomagnetic storm, primarily based on their orientation \citep[]{Wu:2002}, i.e.\ on the presence of southward magnetic fields \citep[]{Tsurutani:1988, Gonzalez:1994} . 

One of the few studies investigating geo-effective shock-sheath characteristics is the work by \citet{Ontiveros:2010}. The authors found that the correlation between the maximum dawn-to-dusk solar wind electric field and the minimum Dst was strong for storms that included a magnetic ejecta but weak for storms due only to a shock-sheath. In contrast, a good correlation was also found for sheaths  when the integral of the dawn-to-dusk field over the crossing time was considered instead. 

There are at least four main ways for a shock-sheath structure to be geo-effective: (i) a shock with a normal close to the ecliptic will compress pre-existing  southward $B_z$, (ii) a shock with a normal with a large inclination to the ecliptic will ``create'' southward $B_z$ in the sheath by deflecting radial interplanetary magnetic fields (IMF) away from the ecliptic, (iii) field line draping around the magnetic ejecta in the back of the sheath creates $B_z$ which will depend on the orientation of the CME and where it is crossed, and (iv) turbulent fields or compressed planar structures may have  southward magnetic fields in the sheath. All of these processes may create strong southward $B_z$ and they were summarized in Figure~2 of \citet{Gonzalez:1994}. A further possibility, which is related to (iii), is that (v) part of the sheath is actually composed of reconnected magnetic field lines ``eroded'' away from the flux rope and maintaining some of its orientation, similar to a scenario discussed by \citet{Dasso:2006} and \citet{Ruffenach:2015}.  Note that causes (i) and (ii) are related to the upstream magnetic field and the shock normal, whereas (iii) and (v) are associated with the orientation of the magnetic ejecta. It is as yet unclear what determines the orientation of turbulent fields and compressed planar structures, i.e. the exact cause of (iv). Note that out-of-ecliptic fields within a given sheath is likely a combination of a number of different mechanisms described above.

In this article, we investigate all shocks whose sheaths resulted in at least a moderate geomagnetic storm over 18.5 years of time covering SC23 and the first half of SC24 (1997-- 06/2015). We aim to determine what makes them different from non-geo-effective shock-sheaths. In section~\ref{sec:study}, we present our results of all geo-effective shock-sheaths, before looking separately at intense and moderate geomagnetic storms. We give examples of different types of geo-effective sheaths in section~\ref{sec:example}. In section~\ref{sec:discussion}, we discuss our results and conclude. 

\section{Study}\label{sec:study}

Statistical results based on identifications by other researchers can give us some insight into the geo-effectiveness of shock-sheaths. However, none of them include all geo-effective shock-sheaths irrespective of whether or not a magnetic ejecta caused another, stronger peak afterwards. Therefore, a more detailed study is required. We focus here on all fast magnetosonic forward shocks, irrespective of their drivers. It is believed that about two thirds of fast forward shocks at 1 AU are driven by CMEs \citep[]{Jian:2006, Jian:2006b}. \citet{Kilpua:2015} also found similar results for the fast forward shocks measured near Earth from 1997 to 2013.  Note that some shocks are driverless at 1~AU \citep[]{Gopalswamy:2009}, i.e.\ they are followed neither by a CME nor by a CIR. \citet{Gopalswamy:2009} argued that many of these driverless shocks are associated with CMEs originating close to a coronal hole. In this study, we distinguish between shocks followed by a magnetic ejecta and shocks not followed by a magnetic ejecta. This latter category is composed of shocks driven by CIRs as well as driverless shocks.

We use the Dst index as the measure of geo-effectiveness and we assign a threshold of $-50$~nT to define a geo-effective event. Within these prescribed constraints, we investigate all geo-effective events due to a shock-sheath structure, i.e.\ both those when the main cause of the  geomagnetic storm is a magnetic ejecta as well as those for which the main cause of the storm is the shock-sheath itself. \citet{Zhang:2008} performed a complementary study but only for dips (defined as a temporary decrease of 15~nT or more of the Dst index) of multiple-dip intense geomagnetic storms during SC23. 

\subsection{Data and Event Selection}
We based our study on: (1) the shock list of \citet{Kilpua:2015}\footnote{\url{http://ipshocks.fi}}, (2) the list of moderate and intense storms of \citet{Yermolaev:2010}, (3) the Dst index data from the world data center for geomagnetism of Kyoto University\footnote{\url{http://wdc.kugi.kyoto-u.ac.jp/index.html}} for SC24,  (4)  the list of CMEs from \citet{Richardson:2010} completed by the list of \citet{Jian:2006}, and, (5) OMNI data\footnote{\url{http://omniweb.gsfc.nasa.gov/ow_min.html}}. The shock parameters, such as fast magnetosonic Mach number, shock speed, compression ratios and shock angle are taken from the IPshocks database. The presence of a CME, its start and end time are taken from the list of \citet{Richardson:2010}, with some exceptions as described below.

We adopted the following procedure: we plotted 1-minute resolution OMNI data for all cases when a shock occurred within 36 hours prior to a moderate or intense geomagnetic storm. We determined whether the shock-sheath itself was geo-effective, irrespective of what occurs after the end of the sheath.  For the end of the sheath (i.e.\ the start of the magnetic ejecta), we usually used the list of CMEs of \citet{Richardson:2010}, although there were instances when we disagreed with their timing. Ultimately, the researchers' judgement was used.
 For the vast majority of the selected cases, the shock-sheath causes a clearly identifiable dip in Dst, part of a single-dip or multiple-dip geomagnetic storm. 

Using the list of \citet{Richardson:2010} as well as our previous list of shocks inside CMEs \citep[]{Lugaz:2015a}, into which we added one event (2003 August 18 at 04:42 UT) that was missed in the initial study, we divided the shocks into three categories: i) shocks within a magnetic ejecta, ii) series of shocks, and iii) shocks propagating into nominal solar wind conditions, referred to as ``normal'' shocks. We require that multiple shocks from series of shocks occur within 24 hours from each other and that our visual inspection confirms that the trailing shock propagates within the sheath of the first shock (and not within the magnetic ejecta).

\begin{figure*}[tb]
\centering
{\includegraphics*[width=3.in]{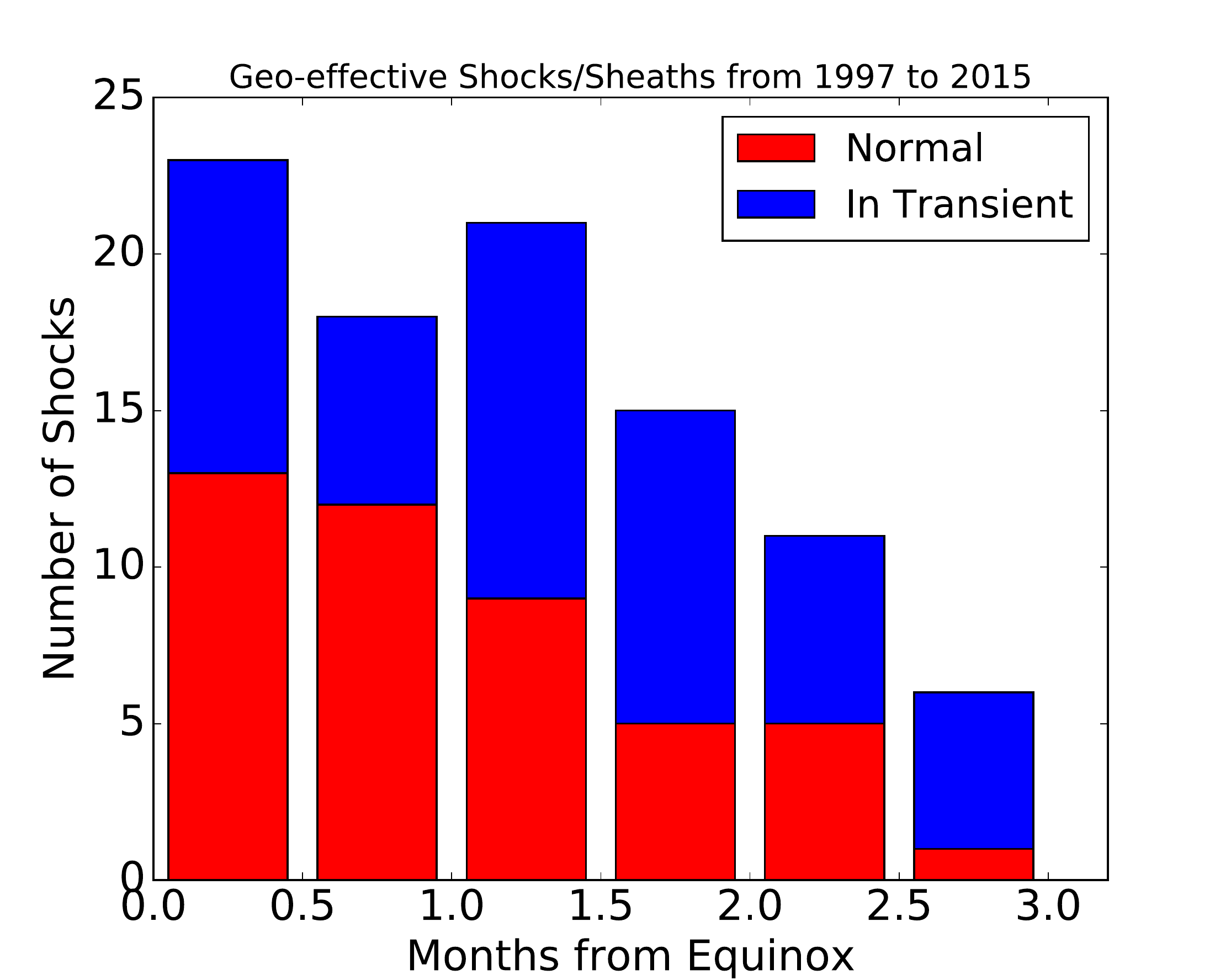}}
\caption{Distribution of the events as a function of the time in month from the closest equinox. Due to the relatively low number of events, Spring and Fall equinoxes are considered together, as is time before and time after the equinox.}
\label{fig:RP}
\end{figure*}

\subsection{General Considerations}
We identified 94 shock-sheaths which caused the Dst index to reach $-50$~nT or lower: 67 cases during SC23 and 27 during SC24. Of these, 38 were intense storms (35 in SC23, 3 in SC24) and 56 were moderate storms (32 in SC23, 24 in SC24). For 78 of these shocks, a magnetic ejecta was identified by \citet{Richardson:2010} as the driver of the shock. 

The left panel of Figure~\ref{fig:dist} shows the yearly distribution of these shocks. The solar cycle dependence is clearly visible, with SCs 23 and 24 separated by a deep solar minimum (2007--2009). The deep solar minimum years 2007-2009 is totally devoid of shock-sheath events which gave rise to moderate or intense geomagnetic storm. In fact, there were only a dozen shocks measured at L1 between 2007 and 2008 (compared to 21 shocks in 1997 alone, for example). The lower number of geo-effective shocks-sheaths during SC24 as compared to SC23 is consistent with results from studies investigating the interplanetary causes \citep[]{Gopalswamy:2015} of geomagnetic storms.  \citet{Gopalswamy:2015} showed that the lower number of storms in SC24 as compared to SC23 is a direct consequence of the lower interplanetary $VB_z$, which in SC24 is 40\% lower than the value in SC23 for magnetic clouds. The associated Dst of magnetic clouds in SC24 is about half that of magnetic clouds in SC23. The effects of the deep solar minimum in 2007--2009 on Earth's magnetosphere was investigated in \citet{Farrugia:2013} and references therein.
The year 2001 near the maximum of SC23 was when most geo-effective shock-sheaths were observed.  A second peak occurred in 2012 at the first maximum of SC24. Overall, as noted above, SC23 was much more active than SC24, with about 6 geo-effective shock-sheaths per year, on average, as compared to 3.5 during SC24. 

 We have also looked at the maximum ap index during the sheath's passage. Since the ap is a 3-hour index (as opposed to the Dst index which has 1 hour resolution) and we only focus on the geo-effects associated with the shock/sheath, for some events, there are some uncertainties on whether the peak of the ap corresponds to the back of the sheath or beginning of the magnetic ejecta. Overall, we found that the median and average peak ap reached during these events was 117 and 131 respectively (corresponding to a Kp of 7- and 7, respectively), with all events corresponding at least to an active geomagnetic period (ap = 32, Kp = 4+), and only 4 events below an ap of 56 (Kp of 5+). This confirms that these shocks/sheaths are associated with, on average, strong geo-effects, whether it is measured by Kp or Dst. A complete description of the interplanetary conditions resulting in the daily average ap index to reach 20 can be found in \citet{Bothmer:2007}.

\subsection{Comparison Between the Two ``Types'' of Shocks} 
From the total of 94 identified geoeffective shocks-sheaths, 45 cases (48\%) were  associated with ``normal'' shocks, 35 cases (37\%) were associated with a shock propagating within an ejecta, and 14 cases (15\%) with series of shocks (for definitions, see Section 2.1). For a few series of shocks, it is difficult to determine whether the second shock propagates within the sheath of the first shock or within a magnetic ejecta that drives the first shock. Overall, both shock propagating within a magnetic ejecta or series of shocks can be thought of as a shock propagating within a previous transient, either the magnetic ejecta or the sheath. For these reasons, we will now treat these two types of shocks as one category, hereafter referred to as ``shocks within transients''.

An estimate of the total number of shocks from 1997 to the end of June 2015 can be obtained by adding the maximum number of fast-mode forward shocks measured by ACE or Wind in each year, resulting in 489 shocks during this time period, the majority (319) being in SC23). As such, the 94 shocks resulting in a geomagnetic storm (moderate or intense) represent about 20\% of the total number of shocks. In other words, one out of every five shocks measured at 1 AU from 1997 to 2015 drove a sheath which resulted in at least a moderate geomagnetic storm. However, as indicated above, it consists of two populations of nearly equal numbers: ``normal'' shocks propagating into nominal solar wind and shocks within transients. Since shocks within transients are rarer, they are in fact more likely to be geo-effective than shocks propagating into nominal solar wind conditions.

Not only are shocks in transients more likely to be geo-effective, but when they are geo-effective, they result, on average, in a stronger geomagnetic storm. The distribution of the peak Dst for the two types of shocks is shown in the right panel of Figure~\ref{fig:dist}. It illustrates how the peak Dst for geo-effective sheaths driven by a shock in a transient is on average lower than that of a sheath driven by a ``normal'' shock (median of $-102$~nT vs.\ $-74$~nT). The two-tailed t-test for the minimum Dst reached for the two samples (``normal'' shocks vs.\ shocks within transients) is 0.0062 (0.0214 without the multiple shocks), which means that there is a statistically significant difference between the two shock types, and therefore, it is justified to discuss them separately. The right panel of Figure~~\ref{fig:dist} also clearly shows that shocks within transients cause the majority of intense geomagnetic storms due to shocks-sheaths.

 The difference between the two types of shocks/sheaths is also statistically significant when looking at the ap or Kp index. Shocks inside transients have a median and average ap of 111 and 149, respectively (corresponding to Kp = 7- and about 7+) as compared to median and average ap of 94 and 112  (corresponding to Kp = 6+  and about 7-), respectively for ``normal' shocks. 

 We also determined the distribution of storms with respect to the time of the equinoxes. \citet{Russell:1973} have shown that geomagnetic activity follows a semiannual variation with the majority of storms occurring close to the equinoxes. We find that this is true for the geo-effective shocks/sheaths as shown in Figure~\ref{fig:RP}. The Russell-McPherron effect is especially pronounced for ``normal'' shocks for which there are four times less geo-effective shocks/sheaths one month from one of the solstices than geo-effective shocks/sheaths one more from one of the equinoxes (4 vs.\ 25). For shocks within transients, the difference is small and might be due to statistical noise (11 vs. 16). Shocks within transients occur primarily in very active solar periods, which appear randomly distributed within the year (March-April 2001 was close to the equinox but October-November 2003 was not).

\subsubsection{Solar Cycle 23}

For solar cycle 23, we can use the list of shocks within CMEs from \citet{Lugaz:2015a} to compare more precisely which proportion of different types of shocks are geo-effective. From 1997 to 2007, there were 30 geo-effective ``normal'' shocks, 27 geo-effective shocks within ejecta, and 10 geo-effective series of shocks. In the same period, there were about 320 shocks in total, so, once again, about 20\% of the shocks drove geo-effective sheaths. Following \citet{Lugaz:2015a} with one additional event found during this study, there were 50 shocks within CMEs during the same time period, therefore about half of these (27/50) resulted in at least a moderate geomagnetic storm. The percentage of ``normal'' shocks which have geo-effective sheaths can only be estimated since the total number of series of shocks is not known, and determining it is beyond the scope of this paper. Two extreme hypotheses are 
(i) that all series of shocks are geo-effective; or that (ii) series of shocks are only as geo-effective as normal shocks. This allows us to estimate that 13 $\pm$ 2\% of the ``normal'' shocks have geo-effective sheaths. 

\begin{figure*}[tb]
\centering
{\includegraphics*[width=6.5in]{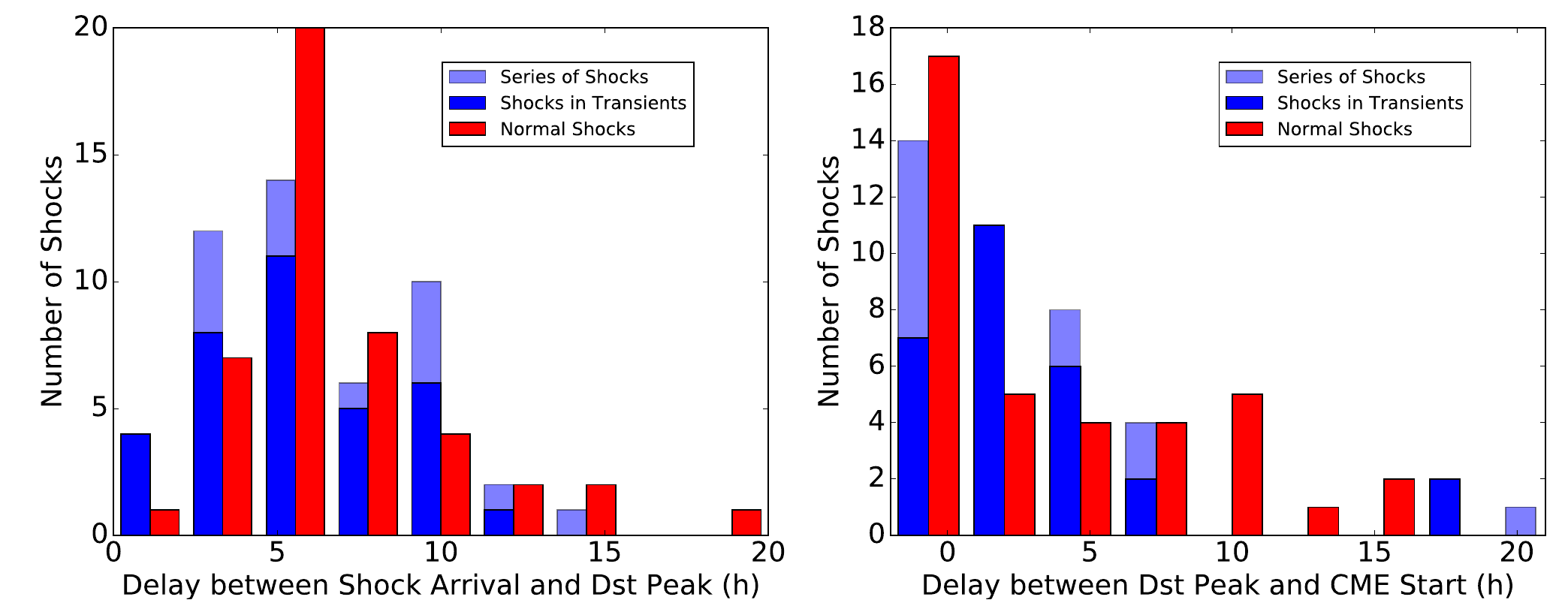}}
\caption{Distribution of geo-effective shocks/sheaths as a function of the delay between the shock arrival and the Dst peak (left) and between the Dst peak and the start of the magnetic ejecta (right).}
\label{fig:dur}
\end{figure*}

\subsubsection{Intense Geomagnetic Storms}
We now focus on intense geomagnetic storms during SC23, since those are the ones that have been investigated in great detail in past studies. As noted earlier, there were 90 intense geomagnetic storms during SC23.  
Nineteen out of the 50 shocks within ejecta caused an intense geomagnetic storms, 6 of the series of shocks and only 10 ``normal'' shocks (out of 200--260 ``normal'' shocks). For SC24, two of the only three intense geomagnetic storms associated with shocks were due to shocks within CMEs (2014/02/19 and 2014/06/22 for shocks within ejecta and 2011/10/24 for a normal shock). 

For validation, we compare our SC23 sample of intense geomagnetic storms caused by shock-sheath structures to what was found by \citet{Zhang:2007} and \citet{Echer:2008}, as summarized in the Appendix. 
Overall, our list agrees with these two previous studies: 26 out of the 30 cases for which \citet{Zhang:2007} identified a shock or a series of shocks as the primary interplanetary cause of an intense geomagnetic storm are also in our list, as do 18 out of the 23 cases from \citet{Echer:2008}.

\subsection{Geomagnetic Storm Timing}


We now calculate the time delay between the shock detection at Wind (or in a few cases at ACE) and the associated minimum of the Dst index. See the left panel of Figure~\ref{fig:dur}. Here, we do not take into account the propagation delay between Wind and the nose of the magnetopause, which is typically in the 20-45 minute range since the studied shocks have speeds in the 500-800 km\,s$^{-1}$ range.

The majority of the shocks cause the Dst index to reach minimum values within six hours (average: 6.67 $\pm$ 3.75 hours; median: 6 hours). This is a relatively short time  keeping in mind that the propagation delay is not taken into account and that shock arrival typically results first in an increase of the Dst index associated with the compression of the magnetosphere (sudden impulse or sudden storm commencement).  For nine cases (9.6\%), however, the peak occurred 12 hours or more after the shock. There is no significant difference in the timing between the different types of shocks. The relatively short time delay (shorter than a typical sheath duration) between the shock being measured at ACE/Wind and the peak Dst attributed to the sheath further confirms that these geomagnetic storms are caused by the sheath field and not by the subsequent driver.
 
As mentioned above, for 78 shocks-sheaths, we are able to associate a magnetic ejecta as the driver of the shock. For these 78 shocks, the average sheath duration is 9.6 $\pm$ 5.2 hours with a median value of 8.7 hours.  We note that the average sheath length at 1~AU has been found to be between 0.083~AU\citep[]{Richardson:2010} and 0.16~AU \citep[]{Jian:2006}, corresponding to a duration of 7--13 hours for a typical 500~km~s$^{-1}$ CME speed. Our sample is therefore fairly typical   in terms of the duration of sheaths.
 Some of the other sixteen shocks are associated with a CIR, whereas for others, there is no clear ejecta after the shock. In some cases of series of shocks or a shock propagating within an ejecta,  we could not identify the start time of the shock driver.

Using the start time of the magnetic ejecta, we estimate that the peak Dst due to the shock-sheath structure was reached on average 3.4 $\pm$ 5.2 hours before the start of the  ejecta. For half of the cases, the peak Dst was reached within $\pm$ 2 hours of the start of the ejecta. 
The right panel of Figure~\ref{fig:dur} shows the distribution of shock-sheaths as a function of the time delay between the peak Dst and the start of the  ejecta.
Both the sheath duration and the location of the Dst peak with respect to the start of the ejecta are statistically the same for ``normal'' shocks as for shocks in transients. 
  
\begin{figure*}[tb]
\centering
\vspace{-0.2cm}
{\includegraphics*[width=6.5in]{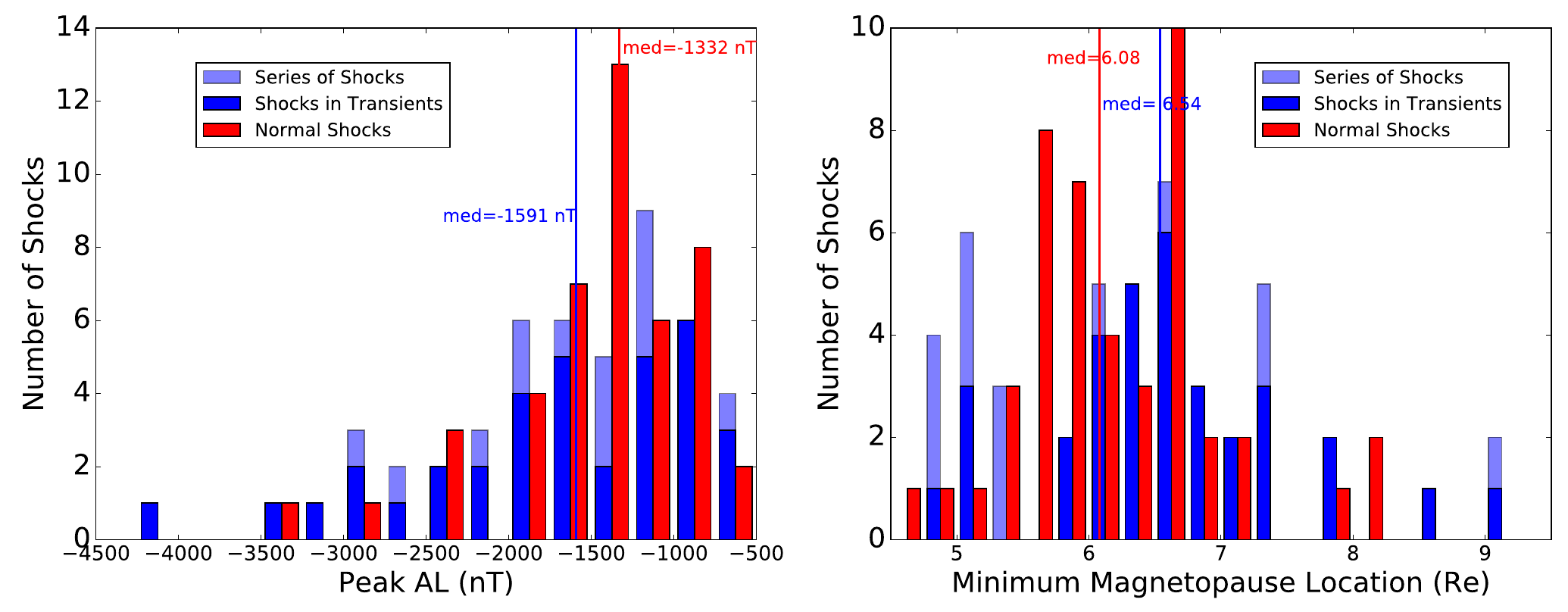}}
\caption{ Distribution of peak AL index (left) and minimum magnetopause location following \citet{Shue:1998} (right) during the sheath passage.}
\label{fig:AL}
\end{figure*}

\subsection{Influence of Shock Parameters}

We found very little correlation between different shock and upstream parameters (compression ratio, shock Mach number, shock speed, upstream values of B and $B_z$) and the minimum Dst. The largest value of cross-correlation is $-0.42$ between the Dst and the shock magnetosonic Mach number for ``normal'' shocks (but no correlation for shocks inside transients), indicating that stronger shocks tend to be more geo-effective, but this is only true for shocks propagating into normal conditions. Overall, the strongest correlation for the full sample is between the upstream magnetic field *strength and the peak Dst with a value of $-0.38$.

\subsection{Substorms}
We now turn to substorms occurring during the sheath passage. Shocks and other solar discontinuities and compressions are known to be able to trigger substorms or intensify pre-occurring substorms \citep[]{Kokubun:1977}, especially when the IMF had a southward component before the shock arrival \citep[]{Zhou:2001,Tsurutani:2003}. \citet{Echer:2011} noted that substorm peak AL-index and the shock strength as measured by the Alfv{\'e}nic Mach number may be correlated. Recently, \citet{Oliveira:2015} performed a dedicated investigation of shocks measured at 1 AU and their effects as measured by the SuperMAG SML index, similar to the AL-index. They only focused on substorms occurring within 2 hours of the shock arrival. Here, we discuss the peak AL-index during the sheath passage at Earth. Some of these substorms are triggered by the shock compression of Earth's magnetosphere, whereas others occur during the main phase of the geomagnetic storms due to the shock-sheath structures.

Figure~\ref{fig:AL} shows the distribution of the peak AL in bins of 250~nT width. There is no statistically significant difference between the ``normal'' shocks and the shocks in transients. We also found no significant correlation between the peak AL index and the shock Mach number or the shock angle with respect to the GSM $x$ direction (i.e.\ a measure of how oblique the shock is). There is however a statistically significant difference in the peak AL of geo-effective shocks-sheaths between SC23 and those from SC24, with medians of $-1561$~nT and $-1013$~nT, respectively. It has been noted before \citep[e.g.\, see][]{Oliveira:2015, Kilpua:2015} the number of shocks was higher in SC23 than in SC24, and we found above that there have been fewer moderate or intense geomagnetic storms due to shocks-sheaths in SC24 than in SC23 (Figure~1). Here, we further find that substorms associated with these shock-sheath events were on average weaker in SC24 as compared to those in SC23. 10 events were associated with ``super-substorms'' when AL index reached below $-2500$~nT \citep[]{Tsurutani:2015}. This corresponds to 8 shocks in transients (6 shocks in ejecta and 2 series of shocks) and 2 shocks in nominal solar wind. This represents 10 of the 37 super-substorms in SC23. 
 Lastly, we found some evidence of a possible difference between shock-sheaths for which the upstream magnetic field was southward (median peak AL  $-1561$~nT, average $-1626$~nT) as compared to those when the upstream magnetic field was northward (median peak AL  $-1253$~nT, average $-1339$~nT). The two-tailed t-test has a p value of 0.0501, which implies that these two groups are different within the 95\% confidence limit (being exactly at the limit of the confidence interval). The difference in the peak AL-index for shocks occurring in northward vs.\ southward IMF is expected, as shocks occurring in pre-existing southward $B_z$ fields have been shown to trigger stronger substorms due to pre-conditioning effects \citep[]{Zhou:2001,Echer:2011}.
  
\subsection{Magnetopause Location}

Shocks and sheaths are a well-known cause of electron dropout events in the outer radiation belt \citep[]{Pulkkinen:2007,Hietala:2014, Kilpua:2015b}. It is beyond the scope of this study to investigate the response of the radiation belt to these shocks-sheaths. However to illustrate their potential effects, we discuss the location of the magnetopause during passage of these shocks-sheaths. \citet{Matsumara:2011} found that earthward motion of the magnetopause is well correlated with subsequent energetic electron losses. We calculated the minimum location of the magnetopause using 1-minute OMNI data and the formula of \citet{Shue:1998} for the 92 events for which OMNI data are available.

The distribution is shown in the right panel of Figure~\ref{fig:AL}. The average minimum magnetopause location during these storms is 6.3 $\pm 0.9$~R$_\mathrm{E}$.  For 70\% of the events (65/92), the magnetopause reached geosynchronous orbit (6.6~$R_\mathrm{E}$). In \citet{Lugaz:2015b}, we discussed how the combination of large dynamic pressures and southward IMF behind shocks within transients can simultaneously compress and erode the magnetosphere, pushing the magnetopause down to below geosynchronous orbit and in some instances, resulting in electron losses through magnetopause shadowing \citep[]{Turner:2012, Alves:2016}. Enhanced wave-particle acceleration and Dst effect \citep[]{Kim:2010} are two other mechanisms which can contribute to energetic electron losses during this type of  events \citep[e.g., see][]{Shprits:2006, Kilpua:2015b}.

We find that the average minimum magnetopause location after series of shocks is 5.9 $\pm 1.3~R_\mathrm{E}$; after a ``normal'' shock, it is 6.2 $\pm 0.75~R_\mathrm{E}$; after a shock inside ejecta, it is 6.6 $\pm 0.9~R_\mathrm{E}$. There is a statistically significant difference between series of shocks and shocks inside ejecta, as well as between shocks inside ejecta and ``normal shocks'', but not between series of shocks and ``normal'' shocks.

\section{Examples of Different Types of Geo-Effective Shock-Sheaths}\label{sec:example}
 In this section, we give examples of shocks followed by geo-effective sheaths, both for ``normal'' shocks as well as for shocks inside transients. The events represent some of the clearest cases rather than typical ones. Whenever possible, we chose recent events, which are less likely to have been studied {\it ad nauseam}. The goal in presenting these case studies is to illustrate events for which the part of the sheath driving the geo-effects can be clearly identified.  In this way, we arrive at a more precise view as to which part of the shock-sheath is responsible for the associated geo-effects when interacting with the magnetosphere.

\subsection{``Normal'' Shocks}

\begin{figure*}[tb]
\centering
{\includegraphics*[width=6.5in]{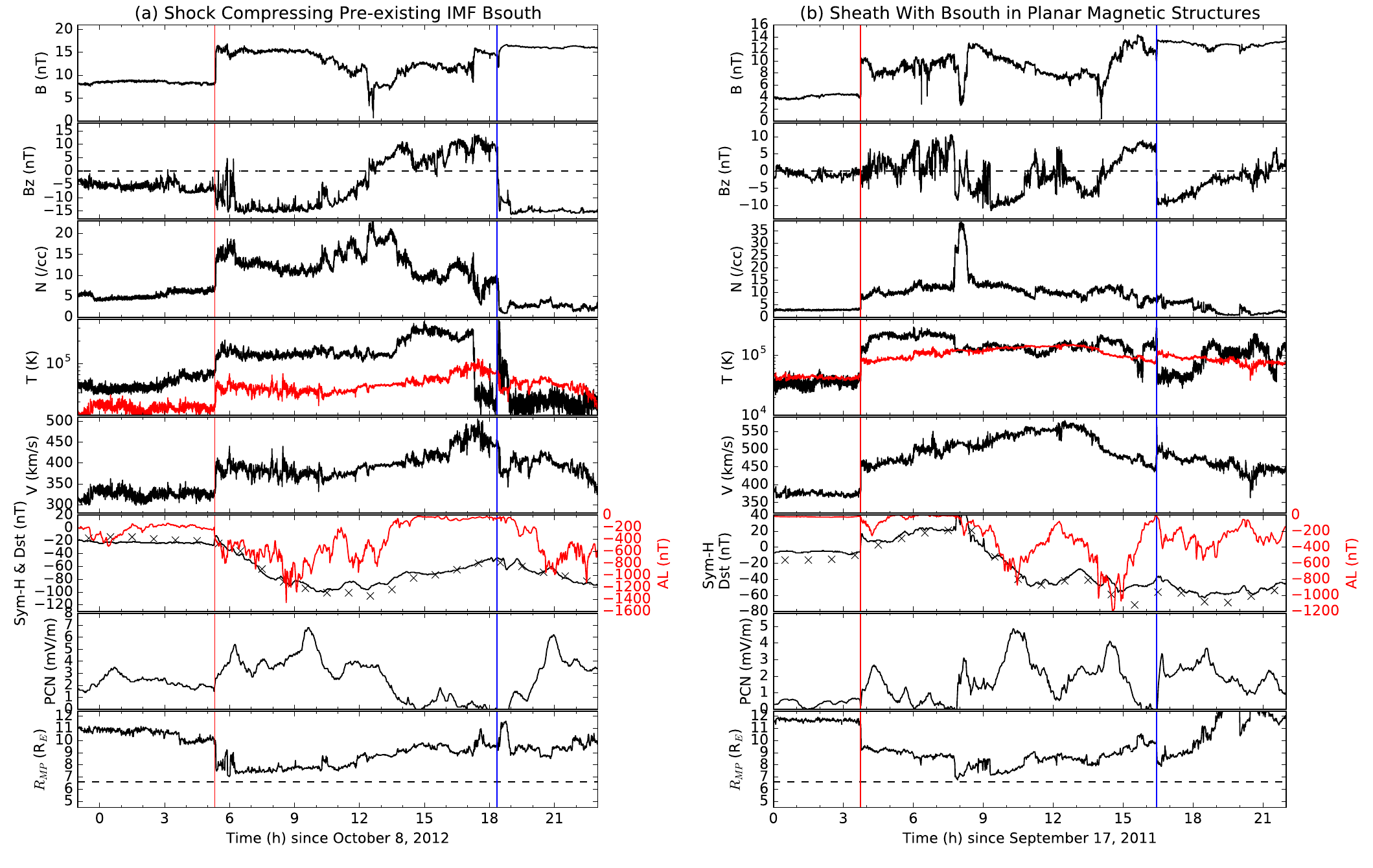}}
\caption{Wind observations of the shock and geo-effective sheaths on October 8, 2012 (Wind shifted by 70 minutes) and September 17, 2011 (Wind shifted by 48 minutes).  The panels show from top to bottom, the magnetic field strength, GSM $B_z$ component, proton density, the proton velocity and temperature (in red, the expected proton temperature), the proton speed,  the Dst index (crosses) and Sym-H index (in red, the AL index) and the PCN index. The CME boundaries as defined by \citet{Richardson:2010} are shown with blue lines, the shock with a red line.}
\label{ex:SH_Bz}
\end{figure*}

\begin{figure*}[t]
\centering
{\includegraphics*[width=6.5in]{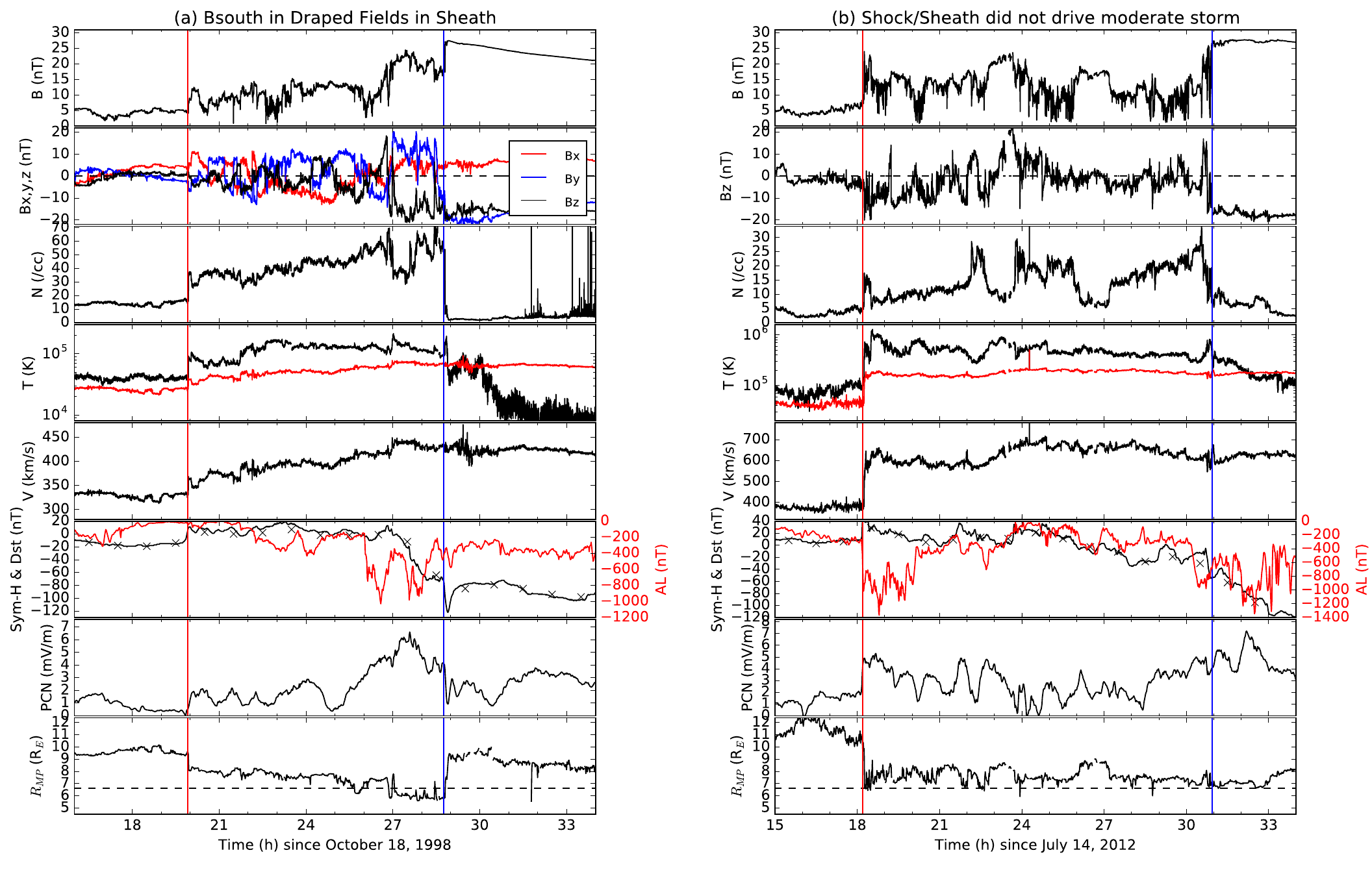}}
\caption{Same as Figure~\ref{ex:SH_Bz} but for the 1998 October 18 (Wind shifted by 27 minutes) and 2012 July 14 shocks (Wind shifted by 36 minutes), the second of which was not geo-effective in terms of Dst index. In the second panel of the 1998 October 18 event, all three components of the magnetic field are plotted.}
\label{ex:SH_CME}
\end{figure*}

Figures~\ref{ex:SH_Bz} and~\ref{ex:SH_CME} show (i) three examples of  ``normal'' shocks that have geo-effective sheaths, and  (ii) one example of  a ``normal'' shock whose sheath did not cause a geomagnetic storm. These events correspond to shocks on 2012 October 8, 2011 September 17, 1998 October 18, and 2012 July 14, respectively. In each Figure, the panels show, from top to bottom, the total magnetic field strength, the southward (GSM) component of the magnetic field, the proton density, the proton temperature (expected temperature following \citet{Lopez:1987} in red), the proton speed, the Sym-H and Dst (crosses) indices, and the AL auroral electrojet index in red, the Polar Cap north (PCN) index determined from the North polar cap station at Thule, Greenland, and the subsolar location of the magnetopause using the formula of \citet{Shue:1998}. All magnetic field and plasma measurements are from Wind, while the geomagnetic indices are from OmniWeb. Wind measurements are time-shifted so as to align the shock arrival with the sudden impulse. The corresponding delays are noted in the figure captions.
The Sym-H index is a 1-minute equivalent to the Dst index and both measure the disturbance in the horizontal component of the geomagnetic field at low latitudes. The AL index reflects the strength of the westward auroral electrojet and is often used to determine substorm activity.  The PCN index quantifies the strength of the cross polar cap convection \citep[]{Troshichev:1996}.

The 2012 October 8 sheath is an example of a shock compressing pre-existing southward magnetic fields in a slow solar wind (Figure 5a).
The shock occurred at 04:12 UT and the peak Dst occurred at 13~UT, i.e.\ about 9 hours after the shock, reaching a value of $-95$~nT(corresponding to a drop of 73~nT with respect to the pre-shock Dst).
The magnetic ejecta starts around 18 UT and is associated with a strong magnetic field, and a drop to low proton densities. The sheath duration was about 13 hours.
 The temperature drop precedes the ejecta by about one hour, and the last hour of the sheath shows signatures of possible erosion \citep[]{Ruffenach:2012},  such as increased proton temperature, and a rotation of the  velocity and magnetic field near the  sharp front boundary of the ejecta.

 Corresponding to a strong southward magnetic field in the ejecta, there was a second drop of Dst (not fully shown) which reached $-105$~nT at 09UT on October 9. This is a classic two-step geomagnetic storm caused by two successive periods of southward $B_z$, the first in the shock-sheath and the second in the magnetic ejecta \citep[]{Kamide:1998}. One hour after the shock, the $B_z$ component of the magnetic field was $\sim -15$~nT, as compared to $-6$~nT before the shock. The shock itself was relatively typical with a compression ratio of 2, a magnetosonic Mach number of 1.9 and a normal with respect to the magnetic field of 74$^\circ$.

As the shock compressed pre-existing southward $B_z$ and the field was northward  towards the end of the sheath, most of the geomagnetic activity occurred during the first half of the sheath. Substorm onset occurred about 3 hours after shock arrival (AL index). This was associated with a strong enhancement in convection (PCN).
From the Sym-H index, we see that  the main phase of the storm started when the shock arrived and lasted seven hours. The PCN index indicates that enhanced convection occurred during the  main storm phase. As $B_z$ turned northward in the second half of the sheath, the
convection decreased, no substorm occurred, and the temporary recovery phase of the geomagnetic storm started.

The 2011 September 17 shock-sheath is an example of a sheath containing geo-effective magnetic fields as part of a planar structure (Figure 5b).  We confirmed this for the period from 07~UT to 12~UT  by performing a minimum variance analysis of the magnetic field \citep[]{Sonnerup:1967}.
The shock occurred at 02:57 UT and the Dst reached $-72$~nT at 16UT (13 hours after the shock). The magnetic ejecta started around 15:40UT. There were two main periods of southward $B_z$ in the sheath: from 07:30UT to 10:10 UT and from 11:45 to 13:20UT. During the first period, $B_z$ was about $-8$~nT. The main phase of the moderate storm was clearly due to the sheath. The sheath duration of 12.5 hours is similar to that in the previous event.
These two periods of southward $B_z$ in the sheath also corresponded to the onset of two substorms associated with two clear and large enhancements in the plasma convection (PCN).cWe recall that convection has a two-source origin, the dayside and the nightside \citep[]{Siscoe:1985,Lockwood:1990}.
Here the nightside (substorm) source was the probably the major contributor.
Before these southward $B_z$ periods, during the first four hours of the sheath, the geomagnetic activity was low. Note that the second substorm was stronger than the first one, even though it was caused by a weaker $B_z$. We have also found strong indications that the whole sheath consisted of Alfv{\'e}nic fluctuations.

The 1998 October 18 shock-sheath (Figure 6a) is an example where draped fields around the magnetic ejecta were associated with strongly southward $B_z$ and drove the geomagnetic activity. The shock occurred at 19:29 UT and the peak Dst at 04UT, i.e., 8.5 hours after the shock, with a value of $-70$~nT. The start of the magnetic ejecta is clear around 04:20~UT, with an extreme drop in density and large drop in temperature, falling below the expected temperature.  This gives a sheath duration of about 9 hours. 
As is clear from the temporal profile of the Sym-H index, the period starting at 02:40 UT with strong southward magnetic fields is the cause of the drop of Dst to moderate storm levels. At the start of the magnetic ejecta, Dst continued to decrease and reached $-112$~nT at 16UT; the drop is approximately monotonic, and the effects of the sheath vs.\ the magnetic ejecta can only be distinguished using the 1-minute Sym-H index. Note that there were a number of short periods of negative $B_z$ during the first half of the sheath but they did not result in any drop of the Dst though they did cause episodic and moderate increases in convection. It is safe to conclude that the draping of the magnetic field around the magnetic ejecta is the main cause of the geo-effective sheath. During the first seven hours of the shock-sheath, i.e.\ before the draped fields, the geomagnetic activity was very quiet.

The draped fields can be considered as part of a planar structure; in fact, \citet{Palmerio:2016} found that about 25\% of the CME sheaths had a planar structure starting just upstream of the magnetic ejecta. For this example, the normal of the planar structure deviates by about $20^\circ$ from the shock normal indicating that its orientation is determined by the magnetic ejecta around which it is draped.

The 2012 July 14 shock-sheath (Figure 6b) is an example of a shock-sheath that was not efficient in energizing the ring current.
This event was chosen from the set of many CME sheaths  that did not result in a moderate geomagnetic storm because $B_z$ was southward before the shock. The shock occurred at 17:39 UT and the peak Dst in the shock-sheath is $-27$~nT at 05~UT on July 15 (about 11.5 hours after the shock). The magnetic ejecta started around 06:20UT and had strong southward $B_z$ magnetic fields. Consequently, the Sym-H and Dst index quickly dropped to the intense geomagnetic storm level, reaching $-118$~nT at 10UT. The sheath duration of 12.5 hours is very similar to the first two events discussed above.
Even in the absence of a moderate geomagnetic storm, the sudden impulse associated with the arrival of the shock when Sym-H increased from 11~nT to 47~nT in 3 minutes coincided with a substorm onset during which the AL-index reached
$-1300$~nT. This substorm is the only notable geomagnetic activity which occurred during the passage of this shock-sheath.
Before shock arrival, the IMF had a relatively strong southward component with an average value of $-2$~nT for the hour before the shock and 5-minute averages reaching $-5$~nT just before the shock. However, the southward field periods within the sheath were generally short and the field magnitude relatively low, and therefore did not result in a geomagnetic storm. 
The substorm onset might have been triggered by the compression of the magnetosphere associated with the shock arrival but the substorm itself cannot be ascribed to the shock-sheath structure.

Although the front of the magnetic ejecta is reminiscent of that of the 1998 October 18 event (Figure 5, left panel),  there was no significant draping in the sheath ahead of the magnetic ejecta. This example serves to emphasize that, while this article focuses on geo-effective shock-sheaths, the large majority of shock-sheaths are not geo-effective, and this even includes events for which the upstream $B_z$ was southward.


\subsection{Shocks Inside Transients}

\begin{figure*}[tb]
\centering
{\includegraphics*[width=6.5in]{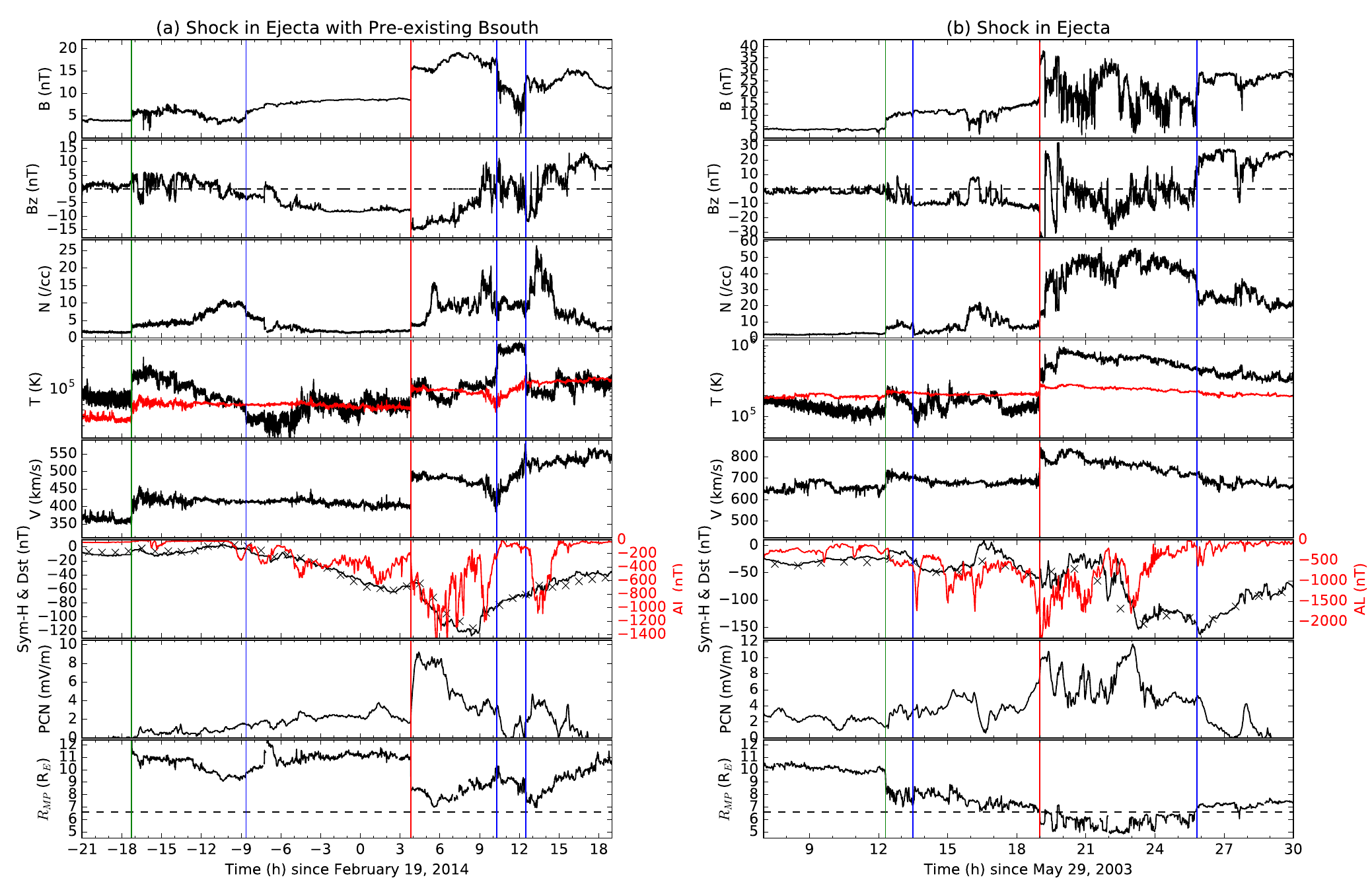}}
\caption{Wind observations of the shock inside transient and geo-effective sheaths on 2014 February 19 (Wind shifted by 39 minutes) and 2003 May 29 (Wind shifted by 29 minutes).  The panels are the same as in the previous Figures. The CME boundaries as defined by \citet{Richardson:2010} are shown with blue lines, the studied shock with a red line, and the previous shock with a green line.}
\label{ex:SHin_Bz}
\end{figure*}

\begin{figure*}[t]
\centering
{\includegraphics*[width=6.5in]{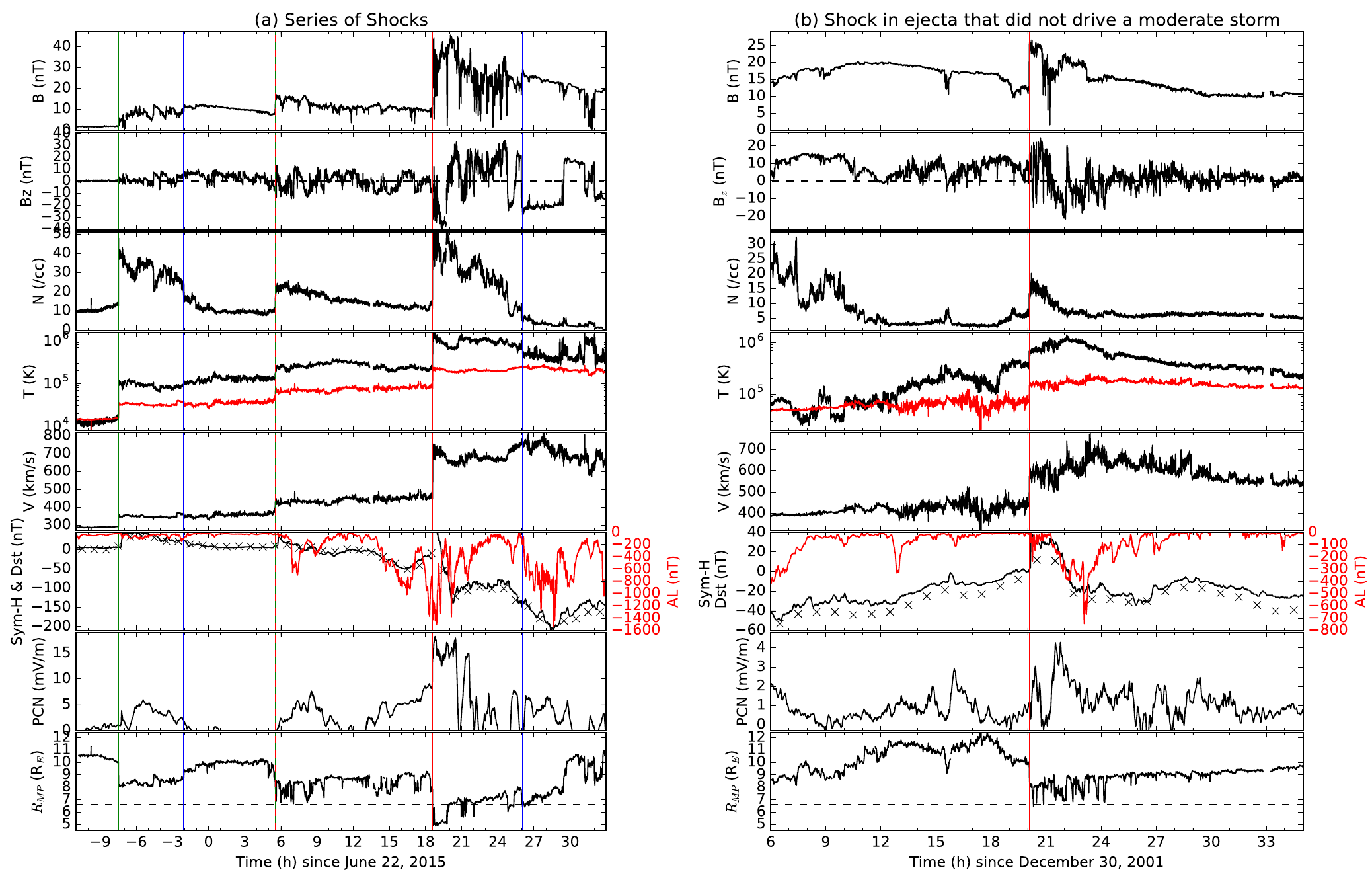}}
\caption{Same as Figure~\ref{ex:SHin_Bz} but for the 2015 June 21-22 series of 3 shocks (Wind shifted by 27 minutes) and for the 2001 December 30 shock (unshifted), which was a shock inside a transient, that did not result in a geo-effective sheath in terms of Dst. For the left panel, the second shock is shown with a red-green line as it resulted in a moderate storm.}
\label{ex:TwoSHs}
\end{figure*}

Figures~\ref{ex:SHin_Bz} and~\ref{ex:TwoSHs} show examples of geo-effective sheaths due to a shock propagating into a previous magnetic ejecta (first two cases), due to a series of shocks (third case) and a non-geo-effective shock propagating within a previous ejecta (last case). These events correspond to the shocks on 2014 February 19, 2003 May 29, 2015 June 22, and 2001 December 30, respectively.

The 2014 February 19 shock-sheath (Figure 7a) provides an example of a shock compressing pre-existing southward magnetic fields that are part of a preceding magnetic ejecta, i.e.\  a specific case related to CME-CME interaction. A first magnetic ejecta started around 15UT on February 18, itself preceded by a very weak shock and a non-geoeffective sheath. A second shock propagated inside this magnetic ejecta, reaching Wind at 03:10 UT. The end of the first magnetic ejecta occurred around 10UT, corresponding to a period lasting about 2 hours with elevated temperature and low magnetic field consistent with compression and reconnection between the two ejecta. All geomagnetic activity was caused by the second shock-sheath.
The peak Dst of $-116$~nT was reached at 09UT, i.e.\ 6 hours after shock arrival and 3 hours before the start of the next CME. The sheath duration (or shocked CME) was about 7 hours.
Before the shock, the Dst reached $-62$~nT at 03~UT consistent with the steady southward magnetic field of $\sim-8$~nT. Following the shock, the southward magnetic increased to $-13$~nT; and associated with the increase of the solar wind velocity from 390 to 480~km~s$^{-1}$, this corresponds to a doubling of the solar wind dawn-to-dusk electric field. The shock arrival was associated with a steep rise in the convection (quadrupling of the PCN index), which coincided with a substorm onset and the growth phase of the storm.
This event is associated with a weak quasi-perpendicular shock with a compression ratio of about 1.7, and a magnetosonic Mach number of 1.9. As is typical of shocks inside CMEs as discussed in \citet{Lugaz:2015a}, this weak shock is still fast with a speed of about 600~km~s$^{-1}$, which is why it caught up with a preceding slower CME.

The 2003 May 29 sheath  (Figure 7b) is another example of a shock propagating through a preceding magnetic ejecta. In this case, the cause of the geo-effective period is harder to pinpoint. There was a first, weak shock at 11:49UT followed by what appears to be a small magnetic ejecta starting around 13UT. A second shock occurred at 18:31UT. This second shock propagated into strong southward magnetic fields, but the sheath was turbulent and the field did not remain southward downstream of the shock. The second magnetic ejecta started around 01UT on May 30 (6.5 hours after the shock) with northward magnetic fields. The peak Dst of $-144$~nT occurred at 00UT, i.e.\ before the start of the second magnetic ejecta. During the sheath, there were three major substorms and large enhancements of the convection. The first substorm appears to be associated with the large $B_z$ oscillation just behind the shock. The third substorm took place during the main phase of the geomagnetic storm.

The 2015 June 22 sheaths are due to a series of three shocks in about  26 hours. These three shocks were of increasing speed.
The first shock at 16:04 UT on June 21 was followed by a magnetic ejecta starting at around 22:30UT. This ejecta is not part of the list of Cane \& Richardson, but its onset appears relatively clear to us due to the smoother magnetic field and lower density. The shock had a compression ratio of about 2.6, a speed of about 310~km\,s$^{-1}$ (the upstream fast magnetosonic speed was of the order of 37~km\,s$^{-1}$). This first shock-sheath and ejecta did not result in any geomagnetic activity.
A second shock at 05:05 UT on June 22 propagated inside this first ejecta and resulted in a peak Dst of $-51$~nT at 17 UT (12 hours after the shock). There was no clear ejecta associated with this second shock. This shock was faster than the first one, with a speed of about 425~km\,s$^{-1}$ and had a compression ratio of 2. There was one substorm associated with the main phase of this first moderate geomagnetic storm.
The third shock at 18:08~UT can therefore be thought of as a shock propagating inside a preceding sheath (or a series of two shocks). It was followed by a magnetic ejecta starting around 02~UT on June 23. This was a fast (765~km\,s$^{-1}$) and strong (compression ratio of 3.6) shock.

The 2001 December 30 shock (Figure 8a) is an example of a shock propagating inside a magnetic ejecta, which did not cause any intensification of the geomagnetic response as measured by the Dst index. Wind measured the passage of a shock-driving magnetic ejecta starting around 00UT on December 30. \citet{Richardson:2010} list 18:00~UT as the end time of this magnetic ejecta, although we find it impossible to identify tits trailing edge due to the presence of the overtaking shock. This overtaking shock crosses Wind at about 20:05 UT. The upstream magnetic field is about 12~nT and the upstream fast magnetosonic speed is in excess of 120~km\,s$^{-1}$, conditions unusual for the unperturbed solar wind. The proton density is relatively low but the proton temperature is higher than expected starting around 12 UT. Overall, the upstream conditions are consistent with the passage of a transient event, either the back part of a magnetic ejecta or the disturbed solar wind in its wake. The shock has a compression ratio of about 2, but the magnetic field remains elevated only for 3 hours after the shock. The IMF $B_z$, initially northward turns southward during this period but does not remain steady. Following a sudden impulse, the Dst reaches $-31$~nT at 02UT on December 31, comparable to its value before the shocks ($-24$~nT at 17UT). There is one moderate substorm (AL reached about $-700$~nT) at the end of the period of enhanced magnetic fields.

The sheath of this third shock compressed pre-existing southward magnetic fields and resulted in a second dip of Dst to $-121$~nT at 21~UT. The magnetic ejecta had strong southward magnetic fields, which resulted in a further decrease of the Dst. The upstream magnetic field had a $B_z$ component of about $-8$~nT; just downstream of the shock, $B_z$ reached $-40$~nT. Two large substorms were triggered by these intense southward fields, and the main phase of the storm occurred at this time. The Sym-H index decreased from 88~nT during the sudden impulse at 18:37UT to $-139$~nT at 20:17~UT at the end of the main phase of the geomagnetic storm. The convection was extremely enhanced during the two hours following the third shock when
substorm activity also took place.

The 2001 December 30 shock  (Figure 8b) is an example of a shock propagating inside a magnetic ejecta, which did not cause any decrease in the Dst index. Wind measured the passage of a shock-driving magnetic ejecta starting around 00UT on December 30. \citet{Richardson:2010} list 18:00~UT as the end time of this magnetic ejecta, although we find it impossible to identify its trailing edge due to the presence of the overtaking shock. This overtaking shock crossed Wind at $\sim$20:05 UT. The upstream magnetic field was about 12~nT and the upstream fast magnetosonic speed was in excess of 120~km\,s$^{-1}$, conditions unusual for the unperturbed solar wind. The proton density was relatively low but the proton temperature was higher than expected starting around 12 UT. Overall, the upstream conditions are consistent with the passage of a transient event, either the back part of a magnetic ejecta or the disturbed solar wind in its wake. The shock had a compression ratio of about 2, but the magnetic field remained elevated for only 3 hours after the shock. The IMF $B_z$, which was initially northward, turned southward during this period but did not remain steady. Following a sudden impulse, the Dst reached $-31$~nT at 02UT on December 31, comparable to its value before the shocks ($-24$~nT at 17UT). There is one moderate substorm (AL reached about $-700$~nT) at the end of the period of enhanced magnetic fields.


\subsection{Series of Shocks}

Since series of shocks were not covered in our previous study \citep[]{Lugaz:2015a}, we summarize their key information in Table~\ref{tab:series} (14 events in total).
Seven cases are clearly shocks propagating into a sheath region of a preceding shock: the two successive shocks occurred within 6.5 hours of each other and as close as 35 minutes (2000 September 17). For all these cases, the second shock was faster than the first one, i.e.\ consistent with its overtaking  the slower shock. However, in several cases the compression ratio of the second shock was weaker than that of the first shock. This is because these overtaking shocks, while fast, are propagating into high upstream speeds and high sound speeds, and are therefore weak in terms of Mach number or compression ratio. This is somewhat similar to shocks propagating inside a magnetic ejecta. The propagation of a fast forward shock inside a sheath of a preceding shock and their subsequent interaction were described in detail in \citet{Lugaz:2005b}. For the five cases when the two shocks occurred within four hours, the magnetic field behind the second shock was of the order of 30~nT, values unusual in sheaths; the total compression in the magnetic field magnitude over these two shocks was typically greater than 5.

For the seven other series of shocks, the two shocks occurred within 20 hours of each other and were eight or more hours apart. In addition, there was no reported magnetic ejecta between the shocks. As the median sheath duration in our study is nine hours, it is likely that these shocks are separated by solar wind plasma or by the leg or draped field around a magnetic ejecta. The solar wind plasma and IMF between the shocks was not that of the typical quiet solar wind, but often had enhanced velocity, density, magnetic field strength, and/or temperature. For three of these series of shocks, the second shock was slower than the leading shock.

\subsection{Statistics}

In addition to the specific examples given above, we attempted to categorize the cause of the geo-effectiveness in the shocks-sheaths for all 94 events, focusing primarily on those 45 cases when the shock-sheath was not associated with a compound event. In 22 out the 45 events, the shock compressing the pre-existing southward IMF was the primary cause of geo-effectiveness. (In three of these cases, draping around the CME also contributed). In five additional cases, draping or eroded magnetic field just upstream of the magnetic ejecta is the main cause of geo-effectiveness. In 17 cases, regions that appear to be planar structures within the sheath were the cause of geo-effectiveness. In one event (2000 November 6), there appeared to be contributions from all three aspects.

In addition, 31 of the shocks inside transients occurred in pre-existing southward magnetic fields. Overall, from the total of 53 events (both including ``normal'' shocks and shocks in transients) for which the shock compressed pre-existing southward magnetic field, the peak Dst occurred 5.4 $\pm 2.6$ hours after the shock, compared to 7.9 $\pm 3.4$ hours after the shock for the other events.

\begin{table*}[bt]
\centering
\scriptsize
\begin{tabular}{|ccccccccc|ccccc|cccc|}
\hline
\multicolumn{9}{|c|}{Shock2} & \multicolumn{5}{c|}{Shock1} &  \multicolumn{4}{c|}{Geoeffects} \\
\hline
Year & Month & Day & HH & MM & V(km\,s$^{-1}$) & $M_\mathrm{ms}$ & $X_N$ & $X_B$ &  $\Delta t$(h) & $\Delta V$ & $M_\mathrm{ms}$ & $X_N$ & $X_B$ & Dst & $\Delta t$(h) & AL & R$_\mathrm{MP}$ \\
\hline
2000 & 9 & 17 & 16 & 46 & 872 & 4.9 & 1.3 & 1.6 & $-0.53$ & $-94$ & 2.8 & 1.5 & 1.8  & $-201$ & 7 & $-1916$ & 5.25 \\
2001 & 11 & 24 & 5 & 51 & 1031 & 4.2 & 5.6 & 5.2 & $-0.95$ & $-410$ & 1.9 & 1.9 & 2.1& $-202$ & 9 & $-2660$ & 4.81 \\
2011 & 8 & 5 & 18 & 40 & 789 & 4.2 & 1.4 & 2.5 & $-1.13$ & $-401$ & 3.1 & 2.3 & 2.8& $-96$ & 5 & $-2014$ & 5.21\\
2001 & 4 & 11 & 16 & 17 & 849 & 1.9 & 2.0 & 1.8 & $-2.13$ & $-137$ & 2.1 & 2.3 & 1.5 & $-215$ & 6 &$-2903$ & 5.31 \\
2001 & 3 & 31 & 1 & 14 & 586 & 5.1 & 2.9 & 5.2 & $-3.57$ & $-276$ & 3.8 & 2 & 1.5& $-156$ & 4 & $-1057$ & 4.83\\
2004 & 4 & 3 & 14 & 51 & 578 & 2.9 & 1.6 & 1.6 & $-4.91$ & $-186$ & 1.4 & 1.6 & 1.6 & $-112$ & 10 & $-1353$ & 6.69\\
2000 & 11 & 26 & 11 & 24 & 579 & 3 & 2.5 & 3.6 & $-6.4$ & $-151$ & 1.8 & 1.8 & 1.6& $-80$ & 14 & $-1111$ & 5.49\\
\hline
2004 & 11 & 7 & 17 & 54 & 733 & 2.6 & 2.2 & 2.3 & $-7.85$ & $-301$ & 3.2 & 2.3 & 3.4& $-117$ & 5 & $-618$ & 5.08\\
2004 & 11 & 9 & 18 & 25 & 808 & 2.6 & 3.0 & 2.2 & $-9.1$ & $-97$ & 3.2 & 4.1 & 2.4& $-214$ & 3 & $-1764$ & 5.11\\
1999 & 9 & 15 & 20 & 8 & 505 & 2.0 & 2.2 & 2.1 & $-12.4$ & 146 & 2.1 & 1.8 & 1.9 & $-67$ & 12 & $-1025$ & 9.08\\
2015 & 6 & 22 & 18 & 8 & 767 & 4.8 & 3.6 & 3.3 & $-13.1$ & $-343$ & 2.0 & 2.0 & 2.0& $-121$ & 3 & $-1376$ & 4.82\\
2015 & 6 & 25 & 18 & 8 & 720 & 2.1 & 1.6 & 1.6 & $-14.9$ & $-24$ & 2.6 & 2.0 & 2.0& $-50$ & 3 & $-1005$ & 7.38\\
2013 & 5 & 25 & 5 & 03 & 401 & 2.2 & 2.2 & 2.1 & $-15.9$ & 126 & 2.4 & 2.9 & 1.9& $-50$ & 9 & $-1260$ & 7.39\\
2006 & 8 & 19 & 10 & 57 & 459 & 1.4 & 1.7 & 1.7 & $-19.1$ & 39 & 2.3 & 1.5 & 1.9& $-78$ & 11 & $-1697$ & 6.19\\
\hline
\end{tabular}
\caption{Series of shocks which resulted in geo-effective sheaths. Peak Dst and AL indices are given in units of nT and the $\Delta t$ corresponds to the time between the second shock and the peak of the Dst index. $R_\mathrm{MP}$ is the minimum (1-min) magnetopause location given in units of $R_\mathrm{E}$ using the formula from \citet{Shue:1998}.}
\label{tab:series}
\end{table*}

\section{Discussion and Conclusions} \label{sec:discussion}

Using {\it in situ} data from 1997 to June 2015, we have identified 94 shocks whose sheaths resulted in at least a moderate geomagnetic storm. We have validated our list by comparing it to those of \citet{Zhang:2007} and \citet{Echer:2008} for the 35 intense storms from SC23, finding general agreement. The largest number of geo-effective shock-sheaths occurred in solar maximum of SC23 (2000--2001) with fewer events in SC24. This is consistent with the lower number of shocks during SC24 as compared to SC23 \citep[]{Oliveira:2015, Kilpua:2015}.

We have found that geo-effective shock-sheaths can best be thought of as two different populations: shock-sheaths associated with CME-CME or CME-CIR interactions (i.e. shocks within transients) and shock-sheaths associated with isolated heliospheric transients (``normal'' shocks). We further distinguish two types of shocks within transients: shocks within magnetic ejecta (35 cases) as studied in \citet{Lugaz:2015a} and series of shocks (14 cases). From past studies, we can estimate that from 1997 to 2006, there were about 200--250 ``normal'' shocks as compared to 50 shocks within magnetic ejecta. Therefore about 50 \% of shocks inside magnetic ejecta are geo-effective as compared to only 13 $\pm$ 2\% of ``normal'' shocks. In addition, shocks within transients result in sheaths which drive, on average, stronger geomagnetic storms, reaching intense level, as compared to moderate level for ``normal'' shocks-sheaths. The difference between the two populations was found to be statistically significant for geomagnetic storms but not for substorms.

To investigate the potential effects of such shock-sheaths on the outer radiation belt, we also calculated the closest subsolar magnetopause location reached during their passage. We found that it is often below geo-synchronous orbit: series of shocks resulted in the most compressed magnetosphere, followed by ``normal'' shocks and then shocks within magnetic ejecta. As each of these shock-sheath structures drove at least a moderate geomagnetic storm, IMF $B_z$ was southward in part of the sheath and there was magnetic erosion of the dayside magnetosphere for all of these shock-sheath structures. The difference in the lowest magnetopause location reached for each of these categories can be understood as the consequence of the dynamic pressure behind each of these shocks. For series of shocks, the sheath plasma has been twice compressed and accelerated, resulting in extremely large dynamic pressures. These sheaths are therefore likely to the most effective in compressing the magnetosphere. On the other hand, shocks inside magnetic ejecta compress the low density plasma typically found inside a magnetic ejecta \citep[]{Burlaga:1981} and result in lower dynamic pressures and consequently lower compression of the magnetosphere. The strong depleting effect of sheath regions has been also highlighted in statistical studies \citep[]{Hietala:2014, Kilpua:2015b}.

We further investigated the  exact cause of the geo-effectiveness of the 94 shock-sheaths, first by studying six examples and, then, by providing statistics on the source of southward $B_z$ in the sheaths. Most shocks inside magnetic ejecta compress pre-existing steady southward magnetic fields, part of the overtaken ejecta. 
We found that, for half the geo-effective sheaths associated with ``normal'' shocks, the shock also compresses pre-existing southward $B_z$. Most other cases appear to be associated with planar structures in the sheath in which the magnetic field was southward. Recently, \citet{Palmerio:2016} studied 95 CME sheaths and found that 85\% of them had planar structures in them, with a median duration of 6 hours. These planar structures are therefore relatively ubiquitous in sheaths and some of them have steady and long-duration (several hours) periods of southward $B_z$.  In fact, \citet{Palmerio:2016} showed that planar parts of the sheath tend to be more geoeffective than non-planar parts. Another good example of such planar structures in a sheath causing a drop of Dst reaching storm levels is the St Patrick 2015 day storm (2015 March 17) as analyzed in \citet{Kataoka:2015}. In a number of cases, these planar structures are at the back of the sheath, just before the start of the magnetic ejecta. In this location out-of-ecliptic magnetic fields and planar structures are likely to result from the draping of magnetic field lines around the magnetic ejecta \citep[]{McComas:1989} and we discussed a specific example of these. Sometimes, these draped fields may also be consistent with eroded CME fields as proposed in \citet{Lavraud:2014}.

We further investigated the  exact cause of the geo-effectiveness of the 94 shock-sheaths, first through case studies and then, through statistics. Most shocks inside magnetic ejecta compress pre-existing steady southward magnetic fields, part of the overtaken ejecta. We found that, for half the geo-effective sheaths associated with ``normal'' shocks, the shock also compresses pre-existing southward $B_z$. Most other cases appear to be associated with planar structures in the sheath in which the magnetic field was southward. Recently, \citet{Palmerio:2016} studied 95 CME sheaths and found that 85\% of them had planar structures in them. These planar structures are therefore relatively ubiquitous in sheaths and some of them contain long periods (several hours) of steady southward $B_z$.  
Another good example of such planar structures in a sheath causing a drop of Dst reaching storm levels is the St Patrick 2015 day storm (2015 March 17) as analyzed in \citet{Kataoka:2015}. In a number of cases, these planar structures are at the back of the sheath, so that they are likely to arise through field line draping around the ejecta \citep[]{McComas:1989}. Sometimes, these draped field may also be consistent with eroded CME fields as proposed in \citet{Lavraud:2014}.

Overall, for 53 (out of 94) cases, $B_z$ was southward before the shock. Hence, we can conclude that the shock compression of this preceding southward fields is the primary driver of geomagnetic storms for the majority of the investigated events. This conclusion can be combined with the fact that shock-sheath structures associated with multiple heliospheric transients are 3--4 times more likely to be geo-effective than shock-sheath structures associated with isolated transients to propose a scheme space weather forecasting for shock-sheath structures. If a shock occurs within a preceding magnetic ejecta, there is a 50\% probability of a moderate geomagnetic storm, and a greater than 30\% probability of an intense geomagnetic storm. This is a similar probability to drive moderate and intense storms as for the passage of a magnetic cloud. If a shock occurs in ``normal'' solar wind conditions, a moderate or intense geomagnetic storm is likely to occur if the IMF had a southward component upstream of the shock. We find little evidence that shocks with a normal direction inclined with respect to the ecliptic are able to create strong and steady southward $B_z$. An inverse study, looking at all instances of shocks occurring in pre-existing southward $B_z$ should be undertaken to validate this conclusion. As such, this is only a proposed scenario. 

While a further dedicated investigation should be undertaken, many of the shock-sheath structures discussed here have been studied recently for their strong effects on radiation belt electrons (for example: 2012 October 1 \citep[]{Baker:2013, Turner:2014a} \& 8 \citep[]{Hudson:2014}, 2013 March 17 \citep[]{Boyd:2014}, 2014 September 12 \citep[]{Alves:2016}, 2015 March 17 \citep[]{Pierrard:2016, Li:2016} and 2015 June 22-25 \citep[]{Baker:2016}). We have found that one of the conditions that make shock-sheath structures geo-effective is a relatively steady southward $B_z$ upstream of the shock due to a preceding transient or  pre-existing southward IMF in the solar wind. The sheath caused by these shocks therefore often combines strong southward $B_z$ and high dynamic pressure. This combination creates ideal conditions for generating large magnetospheric storms and strong losses of energetic electrons from the outer radiation belt (magnetopause shadowing, enhanced Dst effect and wave activity). In some other cases, however, a lower dynamic pressure combined with a southward IMF and high solar wind speed may represent the ideal conditions to accelerate electrons to energy above 1 MeV in the outer radiation belt \citep[]{LiW:2015}.

\begin{acknowledgments}
Wind data used in this study can be downloaded from CDAWeb. The authors acknowledge the use of the OMNIWEB data to obtain the geomagnetic indices and the ICME list of Richardson and Cane. Dst data was obtained from World Data Center for Geomagnetism, Kyoto. We thank the reviewers for their comments which helped clarify some aspects of this manuscript. N.~L. would also like to thank Pr.\ Tsurutani for the initial email exchanges that led to this study. The research for this manuscript was supported by the following grants: NSF AGS-1433213, AGS-1435785, NASA NNX13AH94G, NNX15AB87G, NNXNNX16AO04G and NNX13AP39G, as well as RBSP-ECT funding provided by JHU/APL contract 967399 under NASA's Prime contract NAS5-01072. 
\end{acknowledgments}

\appendix
\section{Comparison of the Interplanetary Causes of Intense Geomagnetic Storms in SC23 to Those Found in Previous Works}

For validation, we compare our sample of intense geomagnetic storms caused by shock-sheath structures in SC23 (35 events) to those found by \citet{Zhang:2007} and \citet{Echer:2008}. In Table~\ref{tab:app}, we list how many events from our list correspond to different classifications from these two past works. For example, our list of 35 shocks/sheaths in SC23 which resulted in an intense geomagnetic storm includes 26 of the 30 intense geomagnetic storms that \citet{Zhang:2007} characterized as being caused by shocks. In performing the comparison, we grouped the different categories from these past studies into (i) storms mainly due to shocks (noted as SH), (ii) storms due mainly to a magnetic ejecta with a secondary contribution from the ejecta-driven shock (noted as (SH) + MC), and, (iii) complex events. A category in parentheses indicates that this interplanetary cause was judged to contribute to the geo-effects, but only secondarily. Note that the events on the same line in both columns are not the same, but the categories are similar. For example, the two ICME(M) from \citet{Zhang:2007} were categorized as ICME and (SH) + MC by \citet{Echer:2008}, whereas the three complex events from \citet{Echer:2008} were categorized as MC, (CME) + PCME-SH and (MC + PCME-SH) + MC by \citet{Zhang:2007}. We found that all eight complex events correspond to a shock propagating within a preceding transient. 
The work by \citet{Echer:2008} and \citet{Zhang:2007} focused on the cause of intense storms and many were multiple-dip storms, where different parts of the ICME drove two or more dips \citep[see][for more details]{Kamide:1998}. As such a category such as (SH) + MC only indicates that the main cause (or the stronger dip) is due to a magnetic ejecta, and we found that in some cases, the shock itself resulted in an intense storm.
Overall, our list agrees with these two previous studies: 26 out of the 30 cases for which \citet{Zhang:2007} identified a shock or a series of shock as the primary interplanetary cause of an intense geomagnetic storm are also in our list, as do 18 out of the 23 cases from \citet{Echer:2008}. \citet{Echer:2008} has a category {\it SH + MC}, for which the relative importance of the shock and the magnetic ejecta is not distinguished. We found that, for 8 out of the 15 events in this category, the shock resulted in an intense geomagnetic storm.

\begin{table*}[bt]
\centering
\begin{tabular}{|c|c|}
\hline
Zhang {\it et al.\ }[2007] & Echer {\it et al.\ }[2008] \\
\hline
{\bf 26/30 SH} &  {\bf 26/38 SH}   \\
{\it 10/12 SH + (MC)} & {\it 17/22 SH} \\
{\it 9/9 PICME-SH} &  {\it 8/15 SH + MC}\\
{\it 7/9 SH(M)} &  {\it 1/1 SH/HCS} \\
\hline
4/29 (SH) + MC  & 3/22 (SH) + MC  \\
\hline
{\bf 4/4 COMPLEX} &  {\bf 4/4 COMPLEX} \\
{\it 2/2 ICME(M), 1/1 CME-CME} & {\it 3/3 COMPLEX} \\
{\it 1/1 (MC + PICME-SH) + MC } & {\it 1/1 PICME-SH}\\
 \hline
& 1/12 CIR \\
1/8 MC &   1/8 ICME \\
\hline
\end{tabular}
\caption{Categorization of the 35 intense geomagnetic storms in SC23 from our list in previous studies by \citet{Zhang:2007} and \citet{Echer:2008}. The second number corresponds to how many of the intense storms were classified under each category by   \citet{Zhang:2007} or \citet{Echer:2008}; the first number corresponds to how many of these events are also in our list. SH: shock, MC: magnetic cloud, ICME: non-MC magnetic ejecta, PICME-SH: shock in previous ICME or MC, (M) indicates multiple (shocks or CMEs), a category in parentheses was considered secondary to the main one.}
\label{tab:app}
\end{table*}

\bibliographystyle{agu}

\end{article}

\end{document}